\documentclass[aps,prd,groupedaddress,showpacs,tighten,floats,nofootinbib]{revtex4}
\usepackage{graphicx}
\usepackage{latexsym}
\def\beq{\begin{equation}}
\def\eeq{\end{equation}}
\def\bey{\begin{eqnarray}}
\def\eey{\end{eqnarray}}

\def\lsim{\mathrel{\raise.3ex\hbox{$<$\kern-.75em\lower1ex\hbox{$\sim$}}}}
\def\gsim{\mathrel{\raise.3ex\hbox{$>$\kern-.75em\lower1ex\hbox{$\sim$}}}}

\begin{document}

\title{Prospects For Detecting Dark Matter With GLAST In Light Of The WMAP Haze}  
\author{Dan Hooper$^{1}$, Gabrijela Zaharijas$^{2}$, Douglas P. Finkbeiner$^{3}$ and Gregory Dobler$^{3}$}
\address{$^{1}$ Fermi National Accelerator Laboratory, Theoretical Astrophysics, Batavia, IL  60510 \\$^{2}$ HEP Division, Argonne National Laboratory, 9700 Cass Ave., Argonne, IL  60439\\$^{3}$ Harvard-Smithsonian Center for Astrophysics, 60 Garden Street, MS51, Cambridge, MA  02138}

\date{\today}

\begin{abstract}

Observations by the WMAP experiment have identified an excess of microwave emission from the center of the Milky Way. It has previously been shown that this ``WMAP Haze'' could be synchrotron emission from relativistic electrons and positrons produced in the annihilations of dark matter particles. In particular, the intensity, spectrum and angular distribution of the WMAP Haze is consistent with an electroweak scale dark matter particle (such as a supersymmetric neutralino or Kaluza-Klein dark matter in models with universal extra dimensions) annihilating with a cross section on the order of $\sigma v \sim 3 \times 10^{-26}$ cm$^3$/s and distributed with a cusped halo profile. No further exotic astrophysical or annihilation boost factors are required. If dark matter annihilations are in fact responsible for the observed Haze, then other annihilation products will also be produced, including gamma rays. In this article, we study the prospects for the GLAST satellite to detect gamma rays from dark matter annihilations in the Galactic Center region in this scenario. We find that by studying {\it only} the inner 0.1$^{\circ}$ around the Galactic Center, GLAST will be able to detect dark matter annihilating to heavy quarks or gauge bosons over astrophysical backgrounds with 5$\sigma$ ($3\sigma$) significance if they are lighter than approximately 320-500 GeV (500-750 GeV). If the angular window is broadened to study the dark matter halo profile's angular extension (while simultaneously reducing the astrophysical backgrounds), WIMPs as heavy as several TeV can be identified by GLAST with high significance. Only if the dark matter particles annihilate mostly to electrons or muons will GLAST be unable to identify the gamma ray spectrum associated with the WMAP Haze.

\end{abstract}
\pacs{95.35.+d;95.55.Ka;95.30.Cq; FERMILAB-PUB-07-461-A}
\maketitle

\section{Introduction}

The Wilkinson Microwave Anisotropy Probe (WMAP) has made the most precise measurements to date of the anisotropies of the cosmic microwave background (CMB) over a wide range of angular scales. These measurements have been invaluable in constraining a wide range of cosmological parameters~\cite{spergel}. Additionally, this data has been used to study standard interstellar medium emission mechanisms, including thermal dust, spinning dust, ionized gas, and synchrotron. After known foregrounds are subtracted from the WMAP data, there remains an unexplained excess of microwave emission within the inner $20^{\circ}$ around the center of the Milky Way, distributed with approximate radial symmetry. This microwave emission has been dubbed the ``WMAP Haze''~\cite{haze1,dobler}.

Although the origin of the WMAP Haze is uncertain, there are compelling reasons to suspect that it may have been generated through the annihilations of dark matter in the form of weakly interacting, massive particles (WIMPs)~\cite{haze2,newhaze}. WIMP annihilations can produce a wide range of possible final states, each of which decay and fragment into a combination of gamma rays, neutrinos, protons, electrons and their antimatter counterparts. A WIMP with an electroweak scale mass will produce relativistic electrons and positrons which, in the presence of the Galactic magnetic field, emit synchrotron photons in the frequency range of WMAP. In previous work, it was shown that the angular distribution of the Haze is consistent with dark matter annihilations with a halo profile scaling as $\rho(r) \propto r^{-1.2}$ in the inner kiloparsecs of our galaxy. The synchrotron spectrum predicted for an electroweak scale WIMP was also found to be consistent with the observations of WMAP. Furthermore, to normalize the dark matter annihilation rate to the observed intensity of the Haze, a dark matter annihilation cross section on the order of $\sim 3 \times 10^{-26}$ cm$^3$/s is required \cite{newhaze}. This is remarkably similar to the value required for an electroweak scale relic to be thermally produced with the observed dark matter abundance. No exotic astrophysical parameters or annihilation boost factors are required to generate the WMAP Haze through this mechanism. 

If the WMAP Haze is in fact generated through the annihilation of dark matter particles in the inner Galaxy, then the intensity and angular distribution of the Haze can be used to estimate the annihilation rate and spatial distribution of dark matter in the inner halo. This can, in turn, be used to estimate the flux of gamma rays and other annihilation products which are expected to accompany the electrons and positrons responsible for the Haze. In this article, we assume that the WMAP Haze results from WIMP annihilation and calculate the corresponding gamma ray spectrum and consider the prospects for the upcoming satellite-based gamma ray telescope GLAST to detect this signal. We find that WIMPs annihilating in the inner 0.1$^{\circ}$ around the Galactic Center to heavy quarks or gauge bosons will be detectable by GLAST with 5$\sigma$ ($3\sigma$) significance if they are lighter than approximately 320-500 GeV (500-750 GeV). If a somewhat wider angular window is observed (out to $\sim$0.5$^{\circ}$), GLAST will be capable of detecting angular extension of the dark matter halo profile. In directions somewhat away from the Galactic Center, the astrophysical background from the TeV gamma ray source observed by HESS is significantly reduced, enabling GLAST to identify the dark matter annihilation radiation associated with the WMAP Haze with high significance. The only case we find in which GLAST would not be expected to identify the dark matter associated with the WMAP Haze is if the dark matter annihilates largely to $e^+ e^-$ or $\mu^+ \mu^-$ final states, resulting in a high ratio of synchrotron to gamma ray radiation.

\section{Gamma Rays, Synchrotron and Inverse Compton Photons From Dark Matter Annihilations In The Inner Galaxy}

To confirm or refute the hypothesis that dark matter annihilations are responsible for the WMAP Haze, some set of complementary observations will certainly be required. WIMP annihilations are expected to deposit energy into a variety of forms, several of which are potentially observable. WIMPs with electroweak scale masses annihilating into the most commonly assumed final states (such as gauge bosons or heavy quarks) transfer most of their energy ($\sim 70\%$) into neutrinos, and smaller fractions into gamma rays (13-20\%), electron-positron pairs (10-13\%) and proton-antiproton pairs (2-5\%). Due to their weak interactions, the resulting neutrinos are likely to be the most difficult to observe, despite their greater luminosity~\cite{Bertone:2004ag} (the exception to this conclusion being the case of WIMP annihilations taking place in the core of the Sun, from which other annihilation products cannot escape~\cite{neutrinosun}). Gamma rays and charged particles produced in dark matter annihilations in the Mikly Way halo, in contrast, each provide promising opportunities for observation.

Gamma rays, once produced, do not interact significantly over Galactic distances. As a result, they can be studied as direct probes of dark matter annihilation by space or ground based gamma ray telescopes, assuming that the annihilation rate is sufficiently high. Charged annihilation products, in contrast, undergo interactions in the interstellar medium which alter their trajectories and lead to significant energy loses. Electrons and positrons, in particular, lose energy through inverse Compton scattering off of starlight, CMB photons and far infrared emission from dust. Interactions with the Galactic magnetic field also cause them to undergo synchrotron energy losses. The relative importance of these processes depends on the energy densities of radiation and magnetic fields. The electron-positron energy loss rate from these two processes is given by~\cite{longair}:
\begin{equation}
-\bigg(\frac{dE_e}{dt}\bigg) = \, \frac{4}{3} \sigma_T \, c \, \bigg(\frac{E_e}{m_e}\bigg)^2 \, ( U_{\rm{rad}} + U_{\rm{mag}}) \approx 8 \times 10^{-16} \, {\rm GeV/s} \, \bigg( \frac{E_e}{1\, {\rm GeV}}\bigg)^2 \, \bigg[\bigg( \frac{U_{\rm rad}}{5 \, {\rm eV/cm}^3} \bigg)    + 0.4 \times \bigg( \frac{B}{10 \, \mu {\rm G}}\bigg)^2\bigg],
\label{loss}
\end{equation}
where $\sigma_T$ is the Thompson scattering cross section, $U_{\rm{rad}}$ is the energy density in radiation fields (starlight, etc.) and $U_{\rm{mag}}$ is the energy density in magnetic fields. In calculating the synchrotron rate, we have averaged over the pitch angle. The energy density of CMB photons is approximately 0.3 eV/cm$^3$. In the solar neighborhood, starlight is roughly twice as prevalent. At small galactrocentric distances, however, the density of starlight is expected to be considerably larger~\cite{isrf}, which is reflected in our nominal estimate of 5 eV/cm$^3$ used in Eq.~\ref{loss}.

The processes of synchrotron radiation and inverse Compton scattering each lead to potentially observable byproducts~\cite{syndm}. Synchrotron photons with microwave frequencies, as we have stated before, may constitute the observed WMAP Haze. The inverse Compton scattering of highly relativistic electrons and positrons with starlight photons, on the other hand, can generate MeV-GeV photons with a total luminosity $(U_{\rm rad}/U_{\rm mag})$ times brighter than the WMAP Haze. 

\begin{figure}
%\begin{center}

\resizebox{9.5cm}{!}{\includegraphics{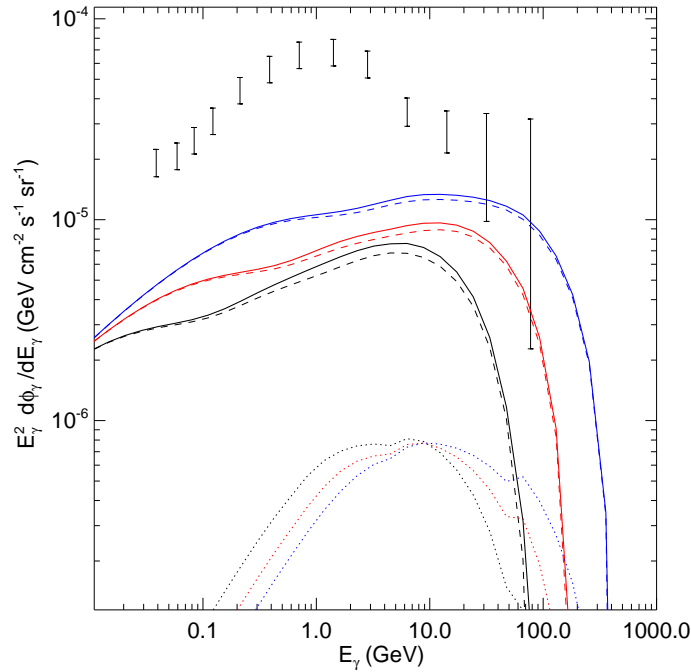}}
\caption{Shown as dashed lines is the gamma ray spectrum from the inverse Compton scattering of relativistic electrons/positrons (required to produce the WMAP Haze) with starlight. In each case, a spectrum of electrons/positrons given by $dN_e/dE_e \propto E_e^{-1}$ (before diffusion) with a spatial distribution proportional to $r^{-2.4}$, where $r$ is the distance from the Galactic Center, was used. The blue, red and black (from top-to-bottom) curves denote spectra which extend to 100, 200 and 500 GeV, respectively. The dotted curves denote the much smaller contribution from Bremsstrahlung emission. The solid lines are the sum of inverse Compton and Bremsstrahlung. For comparison, we also show as error bars EGRET's measurements of the diffuse emission from the inner Galaxy. The lines do not include the gamma ray signal from pion decays produced by proton cosmic rays interacting with the interstellar medium. For reasonable models~\cite{pion}, these gamma rays can account for the measured flux below 1 GeV or so. For this calculation, we used the GALPROP program.}
\label{icplot}
%\end{center}
\end{figure}

In Fig.~\ref{icplot}, we show the spectrum of inverse Compton emission from the inner Galaxy ($l=\pm 30^{\circ}$, $b=\pm5^{\circ}$) associated with the electrons/positrons responsible for the observed WMAP Haze. In this calculation, we have used the starlight spectrum of Porter and Strong~\cite{porterstrong} and an injected spectrum of relativistic electrons/positrons given by $dN_e/dE_e \propto E_e^{-1}$ with a spatial distribution proportional to $r^{-2.4}$, where $r$ is the distance from the Galactic Center. This electron spectrum and distribution were chosen to approximately yield the observed synchrotron Haze. For comparison, we also show the measurements of the diffuse gamma ray spectrum from this region, as measured by EGRET~\cite{egretbars}. Shown as dotted lines is the much smaller and negligible contribution from Bremsstrahlung radiation.

Although the observation of inverse Compton photons from the electrons/positrons responsible for the WMAP Haze could potentially provide a confirmation of the synchrotron nature of the Haze, it would not help to clarify the origin of those relativistic electrons and positrons. To confirm the dark matter origin of these particles, annihilation products independent of the $e^+ e^-$ population need to be identified. The most promising channel in which to make such a confirmation is the gamma ray spectrum produced directly in WIMP annihilations (as opposed to gamma rays from inverse Compton). Unlike inverse Compton photons, these direct gamma rays will originate from the highly concentrated central region of our Galaxy, where the dark matter density is greatest. Furthermore, for typical WIMPs the total luminosity in prompt gamma rays is expected to be comparable to, or (if $U_{\rm rad} \lsim U_{\rm mag}$) somewhat brighter than, the contribution from inverse Compton. A bright source of energetic gamma rays from this angular region, as we will show, is expected if the WMAP Haze is produced through dark matter annihilations.

\section{Observing The Galactic Center With GLAST}

The next generation satellite-based gamma ray telescope GLAST is scheduled to begin its mission later this year and is expected to take data for a minimum of five years. GLAST will have a peak effective area of 11,000 cm$^2$ (or about 60\% of this value when averaged over its field-of-view~\cite{glast}) and a single photon angular resolution of 0.45$^{\circ}$ at 1 GeV, which improves to 0.1$^{\circ}$ at 10 GeV and to 0.035$^{\circ}$ at 100 GeV. In each of these respects, GLAST represents a considerable improvement on its predecessor, EGRET. In contrast to ground based gamma ray telescopes, GLAST is designed to perform an all-sky survey. Emission from the Galactic Center region is expected to be within GLAST's observable field-of-view during 50.8\% of its mission~\cite{glast}.

The Galactic Center has long been considered to be one of the most promising regions of the sky in which to search for gamma rays from dark matter annihilations~\cite{gchist}. The prospects for this depend, however, on a number of factors including the nature of the WIMP, the distribution of dark matter in the region around the Galactic Center, and the presence of any astrophysical backgrounds.

\begin{figure}
%\begin{center}

\resizebox{7.5cm}{!}{\includegraphics{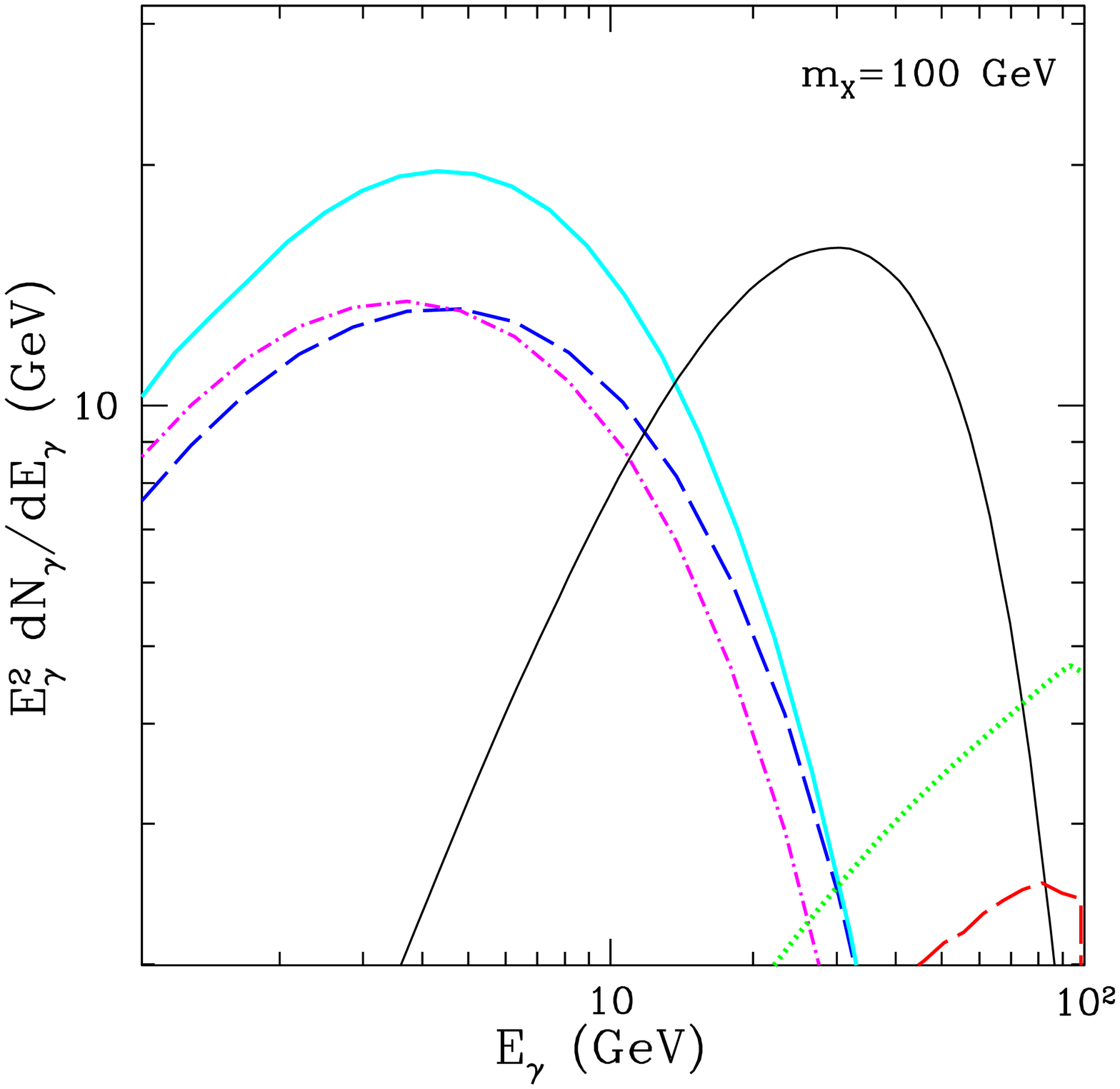}}
\resizebox{7.5cm}{!}{\includegraphics{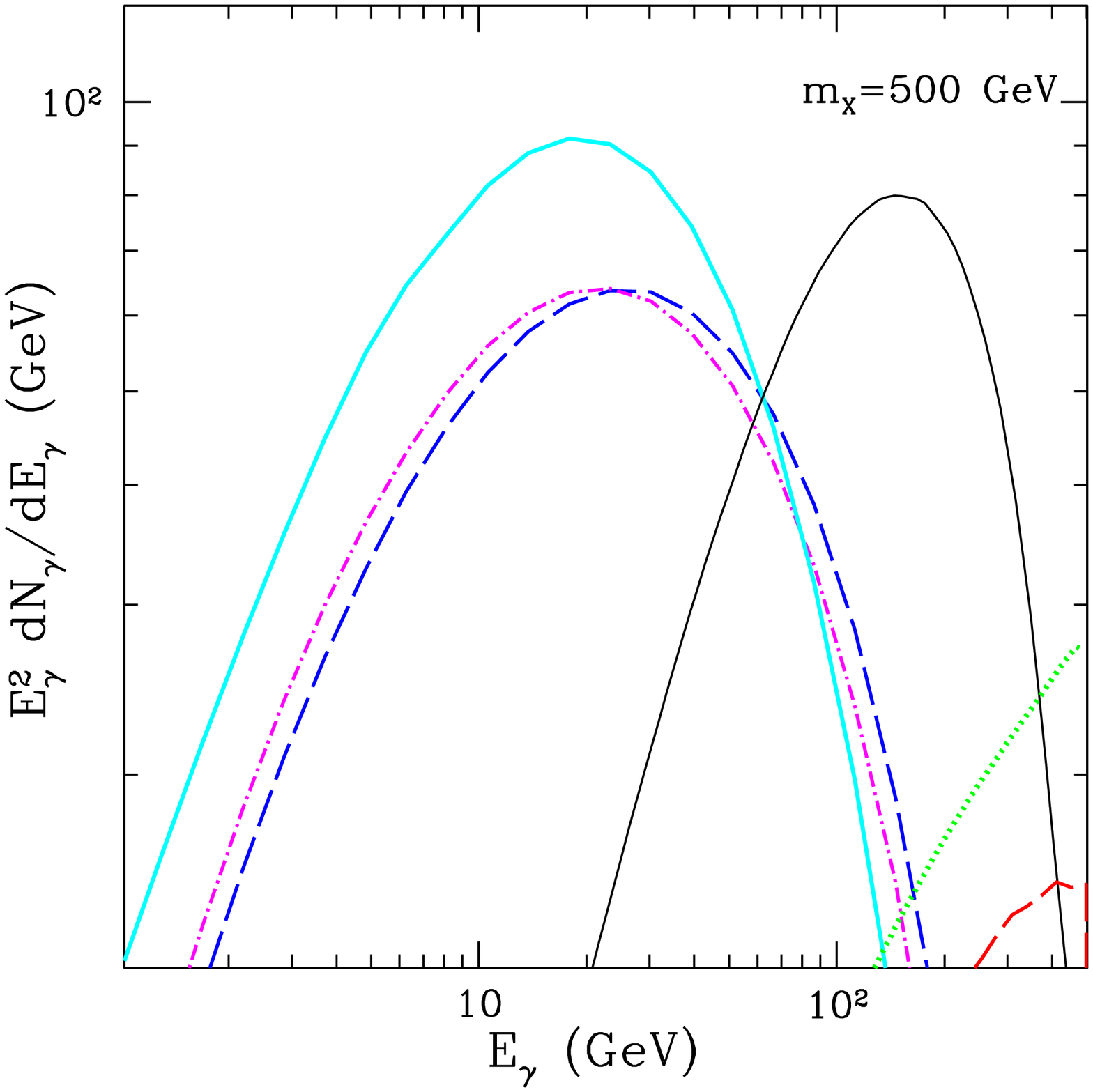}}
\caption{The gamma ray spectrum per WIMP annihilation for a 100 GeV (left) and 500 GeV (right) WIMP. Each curve denotes a different choice of the dominant annihilation mode: $b \bar{b}$ (solid cyan), $ZZ$ (magenta dot-dashed), $W^+ W^-$ (blue dashed), $\tau^+ \tau^-$ (black solid), $e^+ e^-$ (green dotted) and $\mu^+ \mu^-$ (red dashed).}
\label{spectra}
%\end{center}
\end{figure}

The gamma rays produced through dark matter annihilations are described by
\begin{equation}
\Phi_{\gamma}(E_{\gamma},\psi) = \sigma v \frac{dN_{\gamma}}{dE_{\gamma}} \frac{1}{4\pi m^2_X} \int_{\rm{los}} \rho^2(r) dl(\psi) d\psi,
\label{flux1}
\end{equation}
where $\sigma v$ is the WIMP annihilation cross section (multiplied by the relative velocity of the two WIMPs, in the low velocity limit), $m_X$ is the mass of the WIMP, $\psi$ is the angle observed relative to the direction of the Galactic Center, $\rho(r)$ is the dark matter density as a function of distance to the Galactic Center, and the integral is performed over the line-of-sight. $dN_{\gamma}/dE_{\gamma}$ is the gamma ray spectrum generated per WIMP annihilation. The spectrum of photons produced in dark matter annihilations depends on the details of the WIMP being considered. Supersymmetric neutralinos, for example, typically annihilate to final states consisting of heavy fermions ($b \bar{b}$, $t \bar{t}$, $\tau^+ \tau^-$) and gauge or Higgs bosons ($ZZ$, $W^+ W^-$, $W^{\pm}$, $H^{\mp}$, $HA$, $ZH$)~\cite{jungman}. With the exception of the $\tau^+ \tau^-$ channel, each of these annihilation modes result in a very similar spectrum of gamma rays. The gamma ray spectrum from WIMPs annihilating to light leptons can be quite different, however. This can be significant in the case of Kaluza-Klein dark matter in models with one universal extra dimension, for example, in which dark matter particles annihilate significantly to $e^+ e^-$ and $\mu^+ \mu^-$~\cite{kkdm}. In models with two universal extra dimensions, however, annihilations to gauge bosons dominate~\cite{6d}. In Fig.~\ref{spectra} we plot the predicted gamma ray spectrum, per annihilation, for several WIMP annihilation modes.

\begin{figure}
%\begin{center}

\resizebox{10.0cm}{!}{\includegraphics{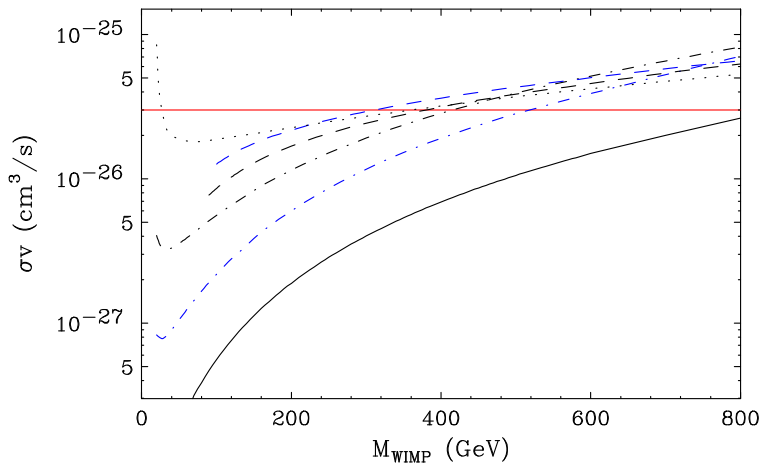}}
\caption{The WIMP annihilation cross section required to produce the intensity of the WMAP Haze. For a wide range of masses and annihilation modes, the cross section required is within a factor of approximately two of the value required of a s-wave thermal relic, $\sigma v \sim 3 \times 10^{-26}$ cm$^3$/s (shown as the horizontal red line). The contours denote the following annihilation modes: $e^+ e^-$(solid), $\mu^+ \mu^-$ (blue dot-dash), $\tau^+ \tau^-$ (dot-dash), $W^+ W^-$ (dashed), $ZZ$ (blue dashed), and $b \bar{b}$ (dotted). (Adapted from Ref.~\cite{newhaze})}
\label{sigma}
%\end{center}
\end{figure}

Assuming that the WMAP Haze is, in fact, the product of dark matter annihilations, and given reasonable assumptions regarding the Galactic magnetic field and radiation fields (see Sec.~\ref{caveats}), we can use the observed intensity to estimate the annihilation rate of dark matter in the region of the Galactic Center. In Ref.~\cite{newhaze}, the required WIMP annihilation cross section was calculated for several possible annihilation modes. In Fig.~\ref{sigma}, we show the required cross sections (adapted from Ref.~\cite{newhaze}) which we will use to normalize the annihilation rate and resulting gamma ray spectrum.

Averaging over a solid angle centered around a direction, $\psi$, we arrive at
\begin{equation}
\Phi_{\gamma}(E_{\gamma}) \approx 2.8 \times 10^{-12} \, {\rm cm}^{-2} \, {\rm s}^{-1} \, \frac{dN_{\gamma}}{dE_{\gamma}} \bigg(\frac{<\sigma v>}{3 \times 10^{-26} \,\rm{cm}^3/\rm{s}}\bigg)  \bigg(\frac{1 \, \rm{TeV}}{m_{\rm{X}}}\bigg)^2 J(\Delta \Omega, \psi) \Delta \Omega,
\label{flux2}
\end{equation}
where $\Delta \Omega$ is the solid angle observed. The quantity $J(\Delta \Omega, \psi)$ depends only on the dark matter distribution, and is the average over the solid angle of the quantity,
\begin{equation}
J(\psi) = \frac{1}{8.5 \, \rm{kpc}} \bigg(\frac{1}{0.3 \, \rm{GeV}/\rm{cm}^3}\bigg)^2 \, \int_{\rm{los}} \rho^2(r(l,\psi)) dl.
\end{equation}
$J(\psi)$ is normalized such that a completely flat halo profile, with a density equal to the value at the solar circle, integrated along the line-of-sight to the Galactic Center would yield a value of one.

In Ref.~\cite{newhaze}, it was shown that the angular distribution of the WMAP Haze is well described by dark matter annihilations with a halo profile approximately given by
\begin{equation}
\rho(r) \approx \frac{\rho_0}{(r/R)^{1.2} \, [1+(r/R)]^2},
\label{1pt2}
\end{equation}
where $\rho_0 \approx 0.3$ GeV/cm$^3$ is the dark matter density in the solar neighborhood and $R \approx 20$ kiloparsecs is the scale radius. This distribution is essentially a Navarro-Frenk-White (NFW) profile~\cite{nfw} with a slightly steeper inner slope: $\rho(r)\propto r^{-1.2}$ rather than $r^{-1}$. This steepening is not unexpected and could naturally result, for example, from the adiabatic contraction of the halo profile~\cite{ac}, or from other effects which are not easily accounted for in N-body simulations. Over a solid angle of $\Delta \Omega = 10^{-5}$ sr (a circle of approximately 0.1$^{\circ}$ radius) around the dynamical center of the Milky Way, this profile leads to a value of $J(\Delta \Omega, \psi=0) \Delta \Omega \approx 1.8$, which is approximately 15 times larger than is found in the NFW case.

Attempting to identify gamma rays from dark matter annihilations taking place near the Galactic Center has been made more challenging by the discovery of a bright, very high-energy gamma ray source in that region, observed by HESS~\cite{hess}, MAGIC~\cite{magic}, WHIPPLE~\cite{whipple} and CANGAROO-II~\cite{cangaroo}. This source appears to be coincident with the dynamical center of the Milky Way (Sgr A$^*$) and has no detectable angular extension (less than 1.2 arcminutes). Its spectrum is well described by a power-law, $dN_{\gamma}/dE_{\gamma} \propto E_{\gamma}^{-\alpha}$, where $\alpha=2.25 \pm 0.04 (\rm{stat}) \pm 0.10 (\rm{syst})$ over the range of 160 GeV to 20 TeV. Although speculations were initially made that this source could be the product of annihilations of very heavy ($\gsim 10$ TeV) dark matter particles~\cite{actdark}, this now appears to be very unlikely. The source of these gamma rays is more likely an astrophysical accelerator associated with our Galaxy's central supermassive black hole~\cite{hessastro}. This gamma ray source represents a formidable background for GLAST and other experiments searching for dark matter annihilation radiation from the Galactic Center region~\cite{gabi}. By studying the gamma rays from directions somewhat away from the Galactic Center, however, it is possible to dramatically reduce the background from the HESS source. For the halo profile shape favored by the angular distribution of the WMAP Haze, there is a sufficient degree of extension to be observed by GLAST.

\begin{figure}
\resizebox{8.5cm}{!}{\includegraphics{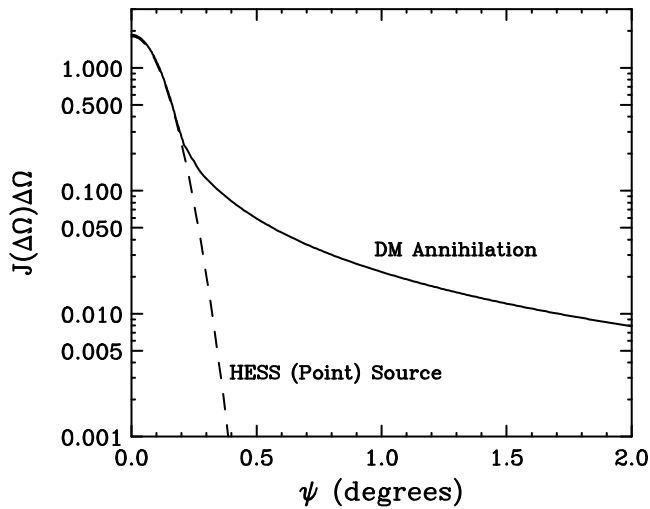}}
\caption{The flux of gamma rays seen by GLAST as a function of the angle observed from the Galactic Center. The dashed line denotes the case of a point source(the astrophysical background), smeared out by the point spread function of GLAST (which we have modeled as a gaussian with a 1$\sigma$ width of 0.1$^{\circ}$). The solid curve is the intensity of dark matter annihilation radiation using the halo profile of Eq.~\ref{1pt2}. The dashed curve has been normalized to match the solid curve at small angles. At angles beyond 0.2$^{\circ}$, the ratio of signal to background is enhanced.}
\label{angle}
%\end{center}
\end{figure}

In Fig.~\ref{angle}, we plot the relative gamma ray intensity as a function of the angle away from the Galactic Center observed by GLAST. Here, we have treated the point spread function of GLAST as a gaussian of width $\sigma=0.1^{\circ}$, which is the approximate width at 10 GeV. The dashed and solid lines denote the cases of a point source (the HESS background) and a distribution corresponding to dark matter distributed with the profile of Eq.~\ref{1pt2}, respectively. Within approximately 0.2$^{\circ}$ of the Galactic Center, the behavior is dominated by GLAST's point spread function, and angular information cannot be used to distinguish the astrophysical point source background from extended dark matter annihilation radiation. At larger angles, however, the extension of the dark matter halo profile becomes important. By observing at angles beyond 0.2$^{\circ}$, the ratio of signal to background is considerably enhanced.

%In Fig.~\ref{angle2}, we show results similar to those of Fig.~\ref{gcspec}, but comparing the spectrum in the angular ranges of $0-0.1^{\circ}$ to that between $0.3^{\circ}$ and $0.5^{\circ}$ from the Galactic Center. Although the overall flux from dark matter annihilation radiation is smaller at larger angles than from the Galactic Center, the background falls off more quickly, making a clear identification far easier. Even for a multi-TeV WIMP, the signal from dark matter annihilations can be detected with high statistical significance by GLAST.

\begin{figure}
%\begin{center}

\resizebox{7.5cm}{!}{\includegraphics{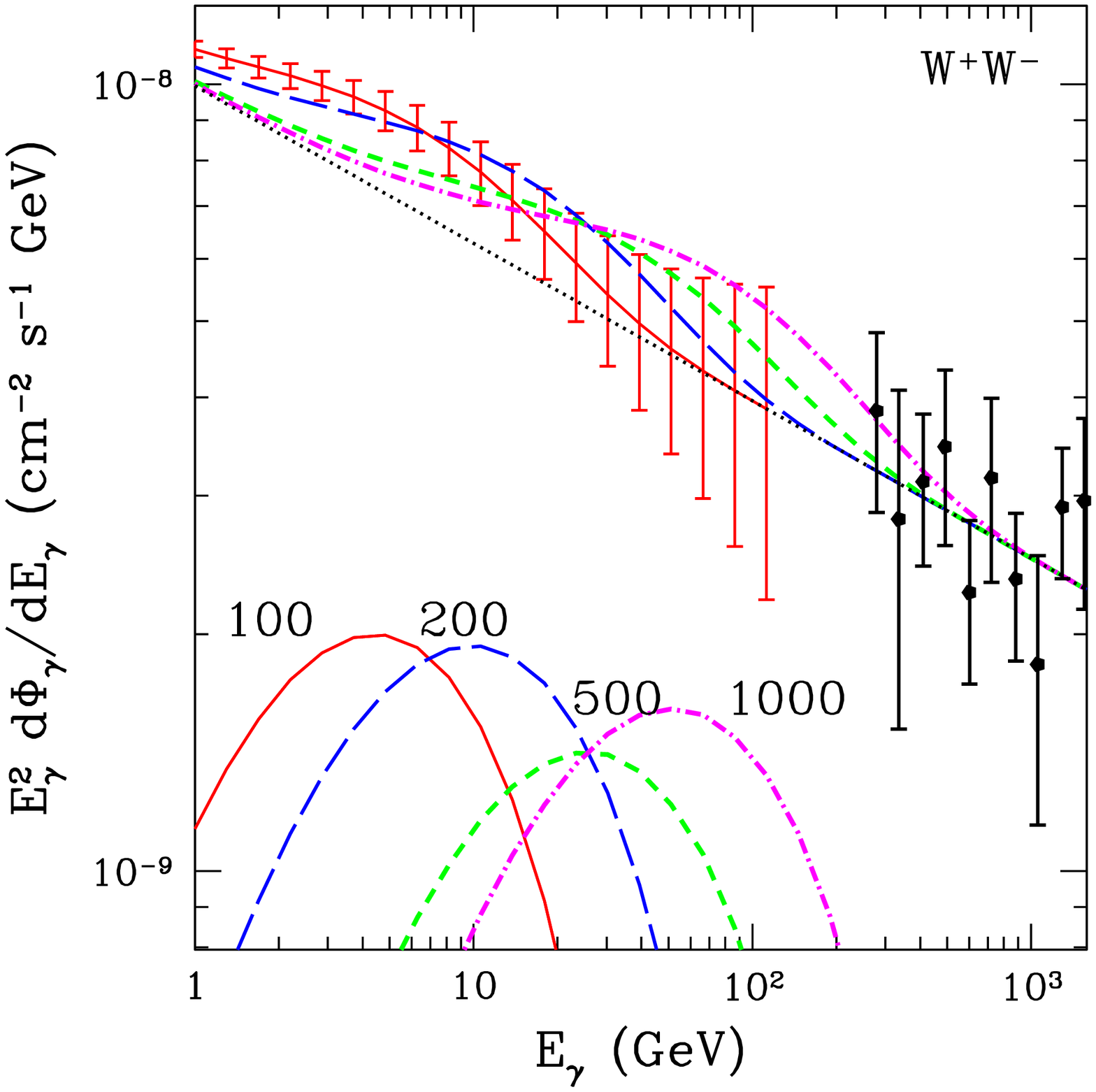}}
\resizebox{7.5cm}{!}{\includegraphics{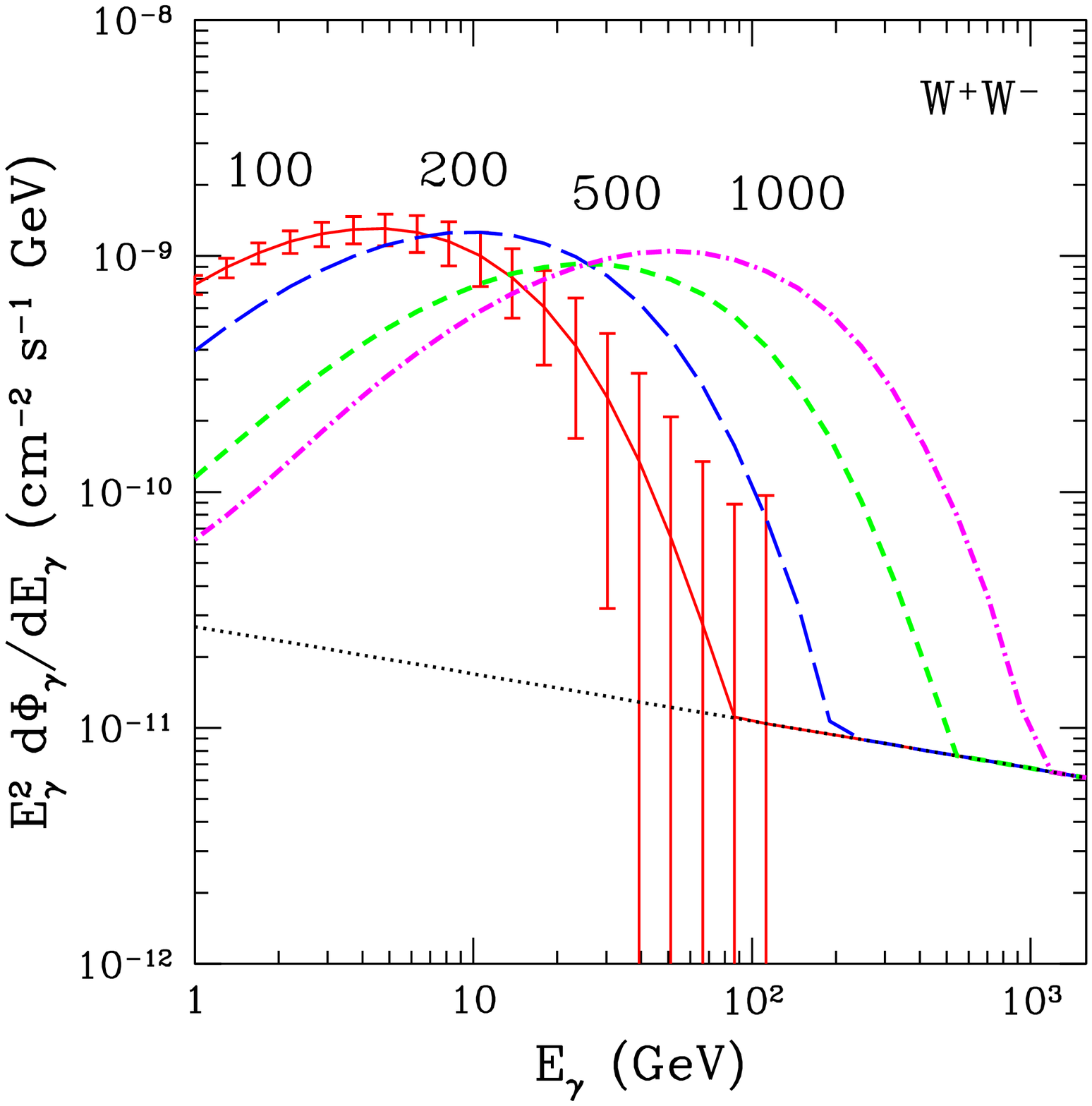}} \\
\resizebox{7.5cm}{!}{\includegraphics{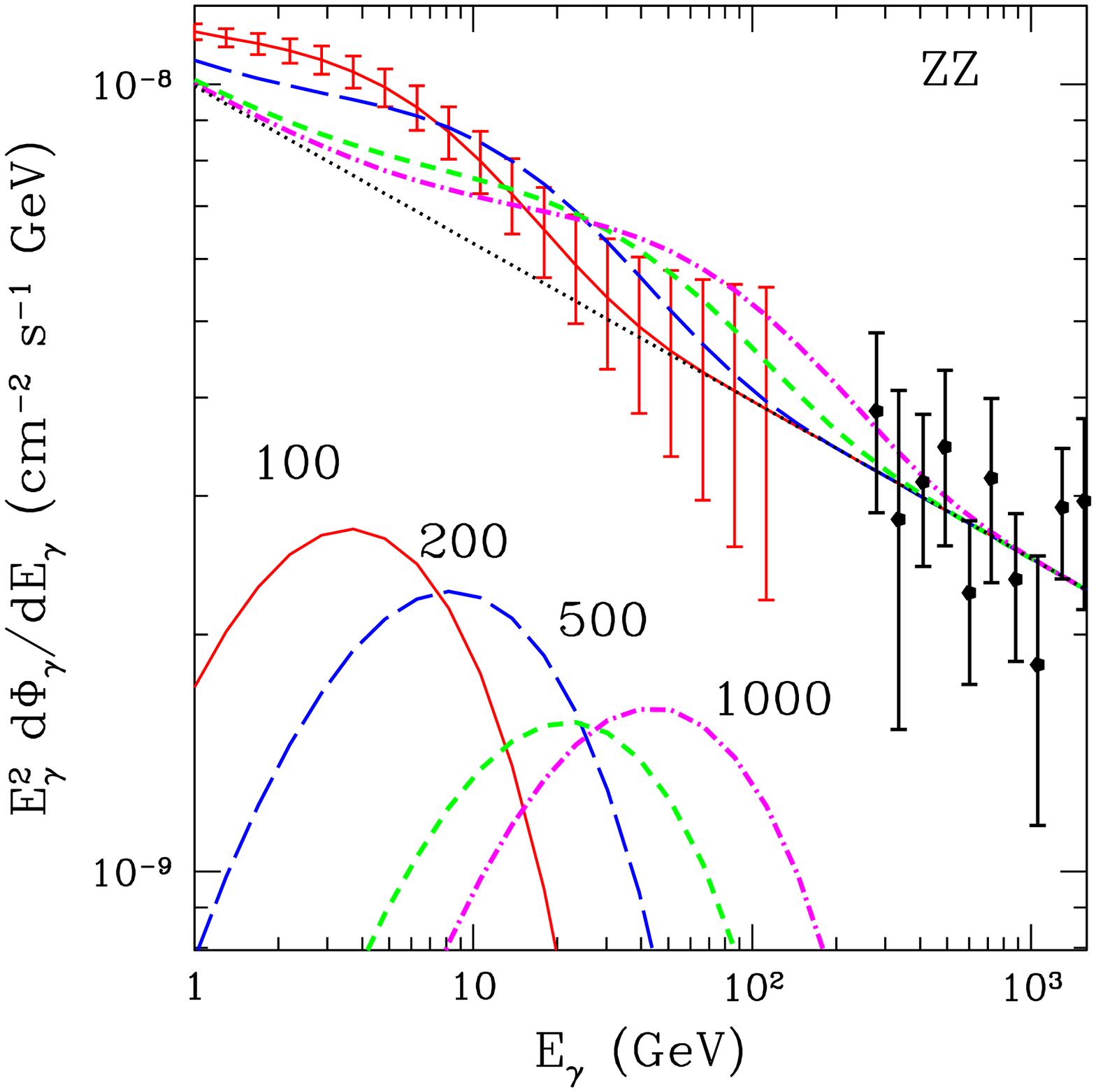}}
\resizebox{7.5cm}{!}{\includegraphics{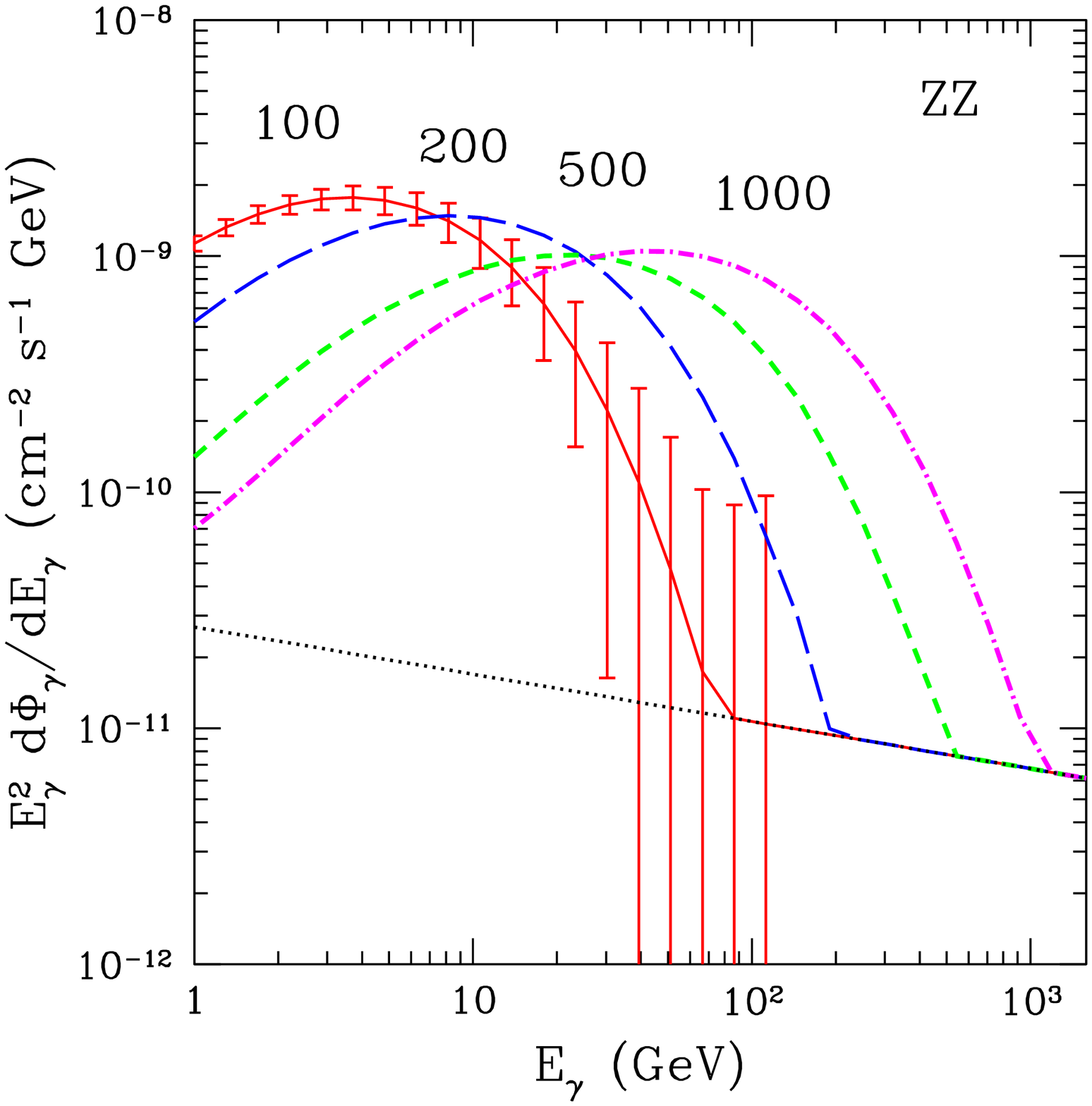}} \\
\resizebox{7.5cm}{!}{\includegraphics{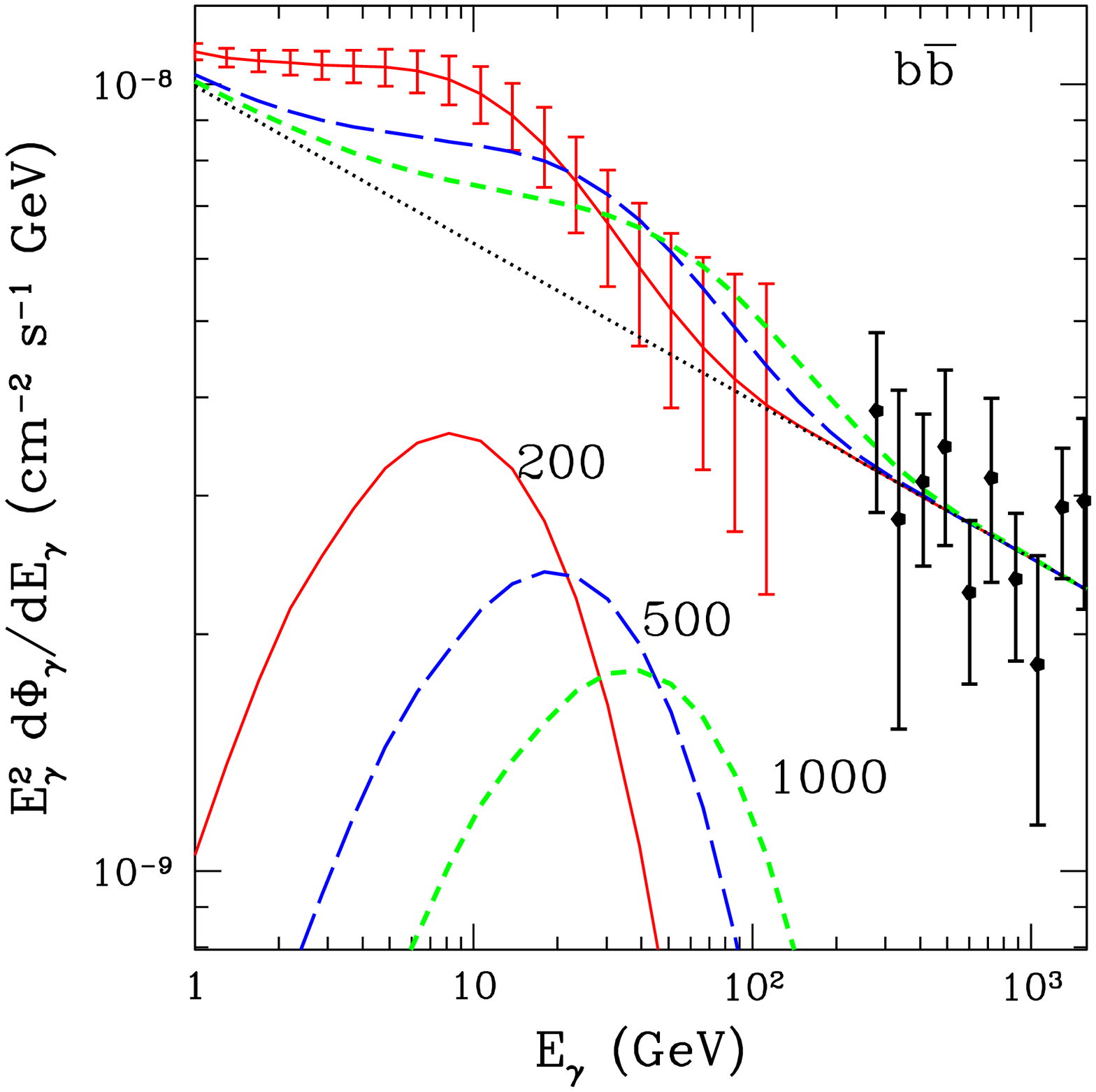}}
\resizebox{7.5cm}{!}{\includegraphics{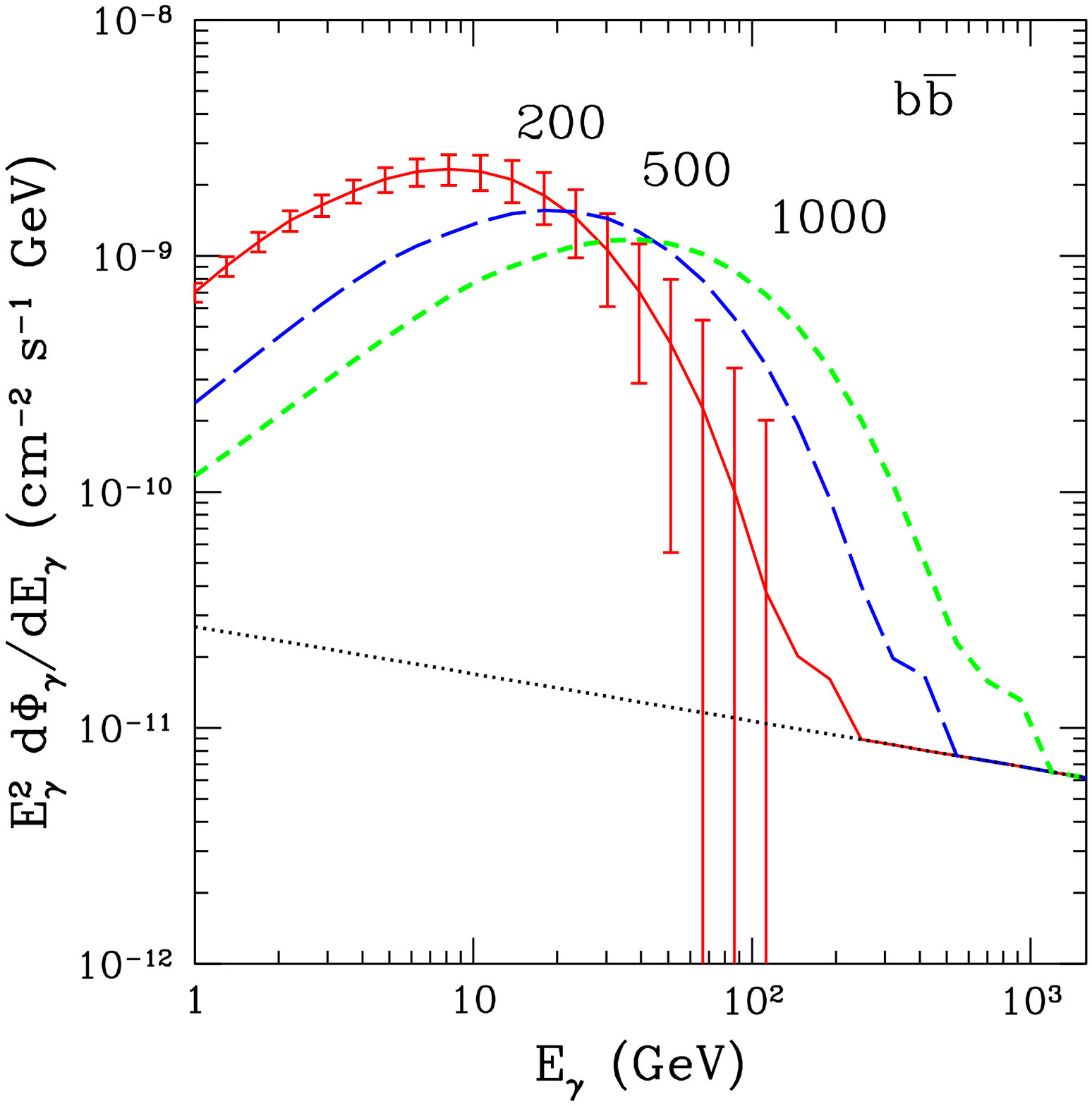}}
\caption{The spectrum of gamma rays from WIMP annihilations in the Galactic Center region, normalized to produce the observed intensity of the WMAP Haze. In the left frames, the flux is averaged over a solid angle of $10^{-5}$ sr (a circle of approximately 0.1$^{\circ}$ radius) centered around the Galactic Center. In the right frames, the result is from angles between 0.3 and 0.5$^{\circ}$ from the Galactic Center. Each frame represents annihilations to a different set of final state particles, and for several WIMP masses (labeled in GeV). We show both the spectra from dark matter annihilations alone, and those spectra plus the background extrapolated from the observed HESS source. Errors bars projected for GLAST are for the case of the lightest WIMP mass shown. The error bars and points shown above 200 GeV in the left frames are from the HESS telescope~\cite{hess}. For the three annihilation modes shown, GLAST should be capable of identifying the component from dark matter annihilations.}
\label{gcspec}
%\end{center}
\end{figure}

\begin{figure}
%\begin{center}

\resizebox{7.5cm}{!}{\includegraphics{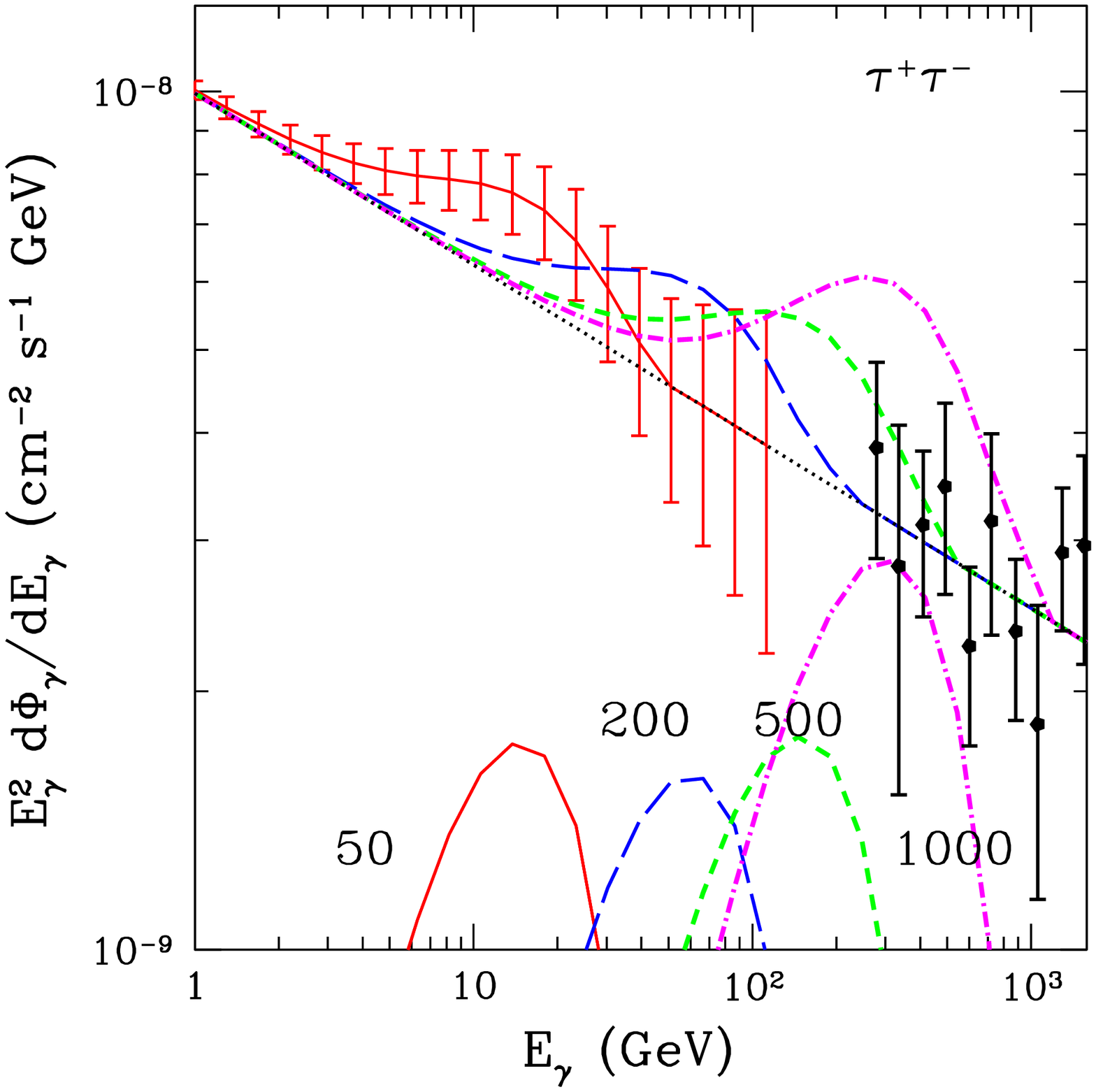}}
\resizebox{7.5cm}{!}{\includegraphics{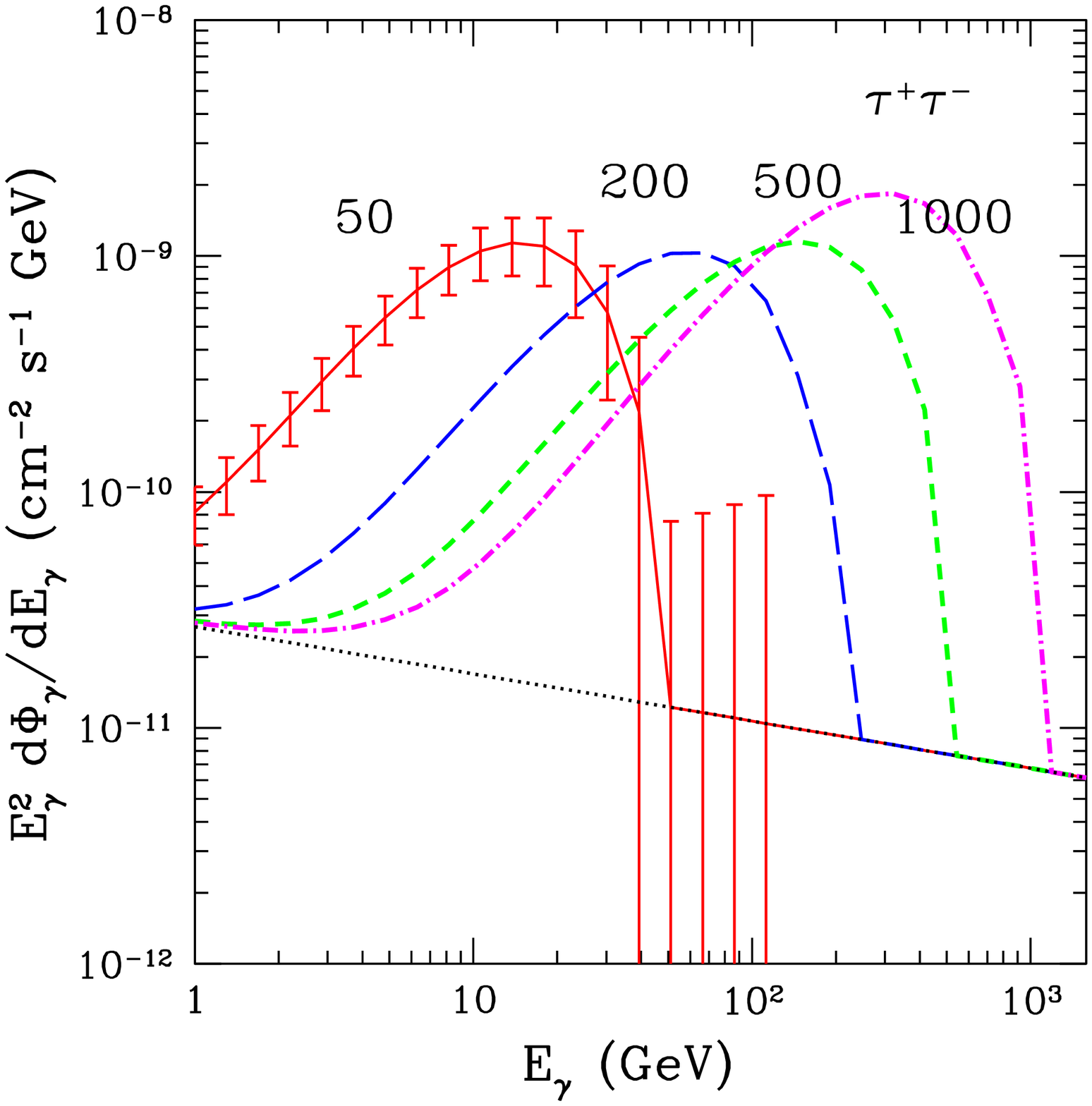}} \\
\resizebox{7.5cm}{!}{\includegraphics{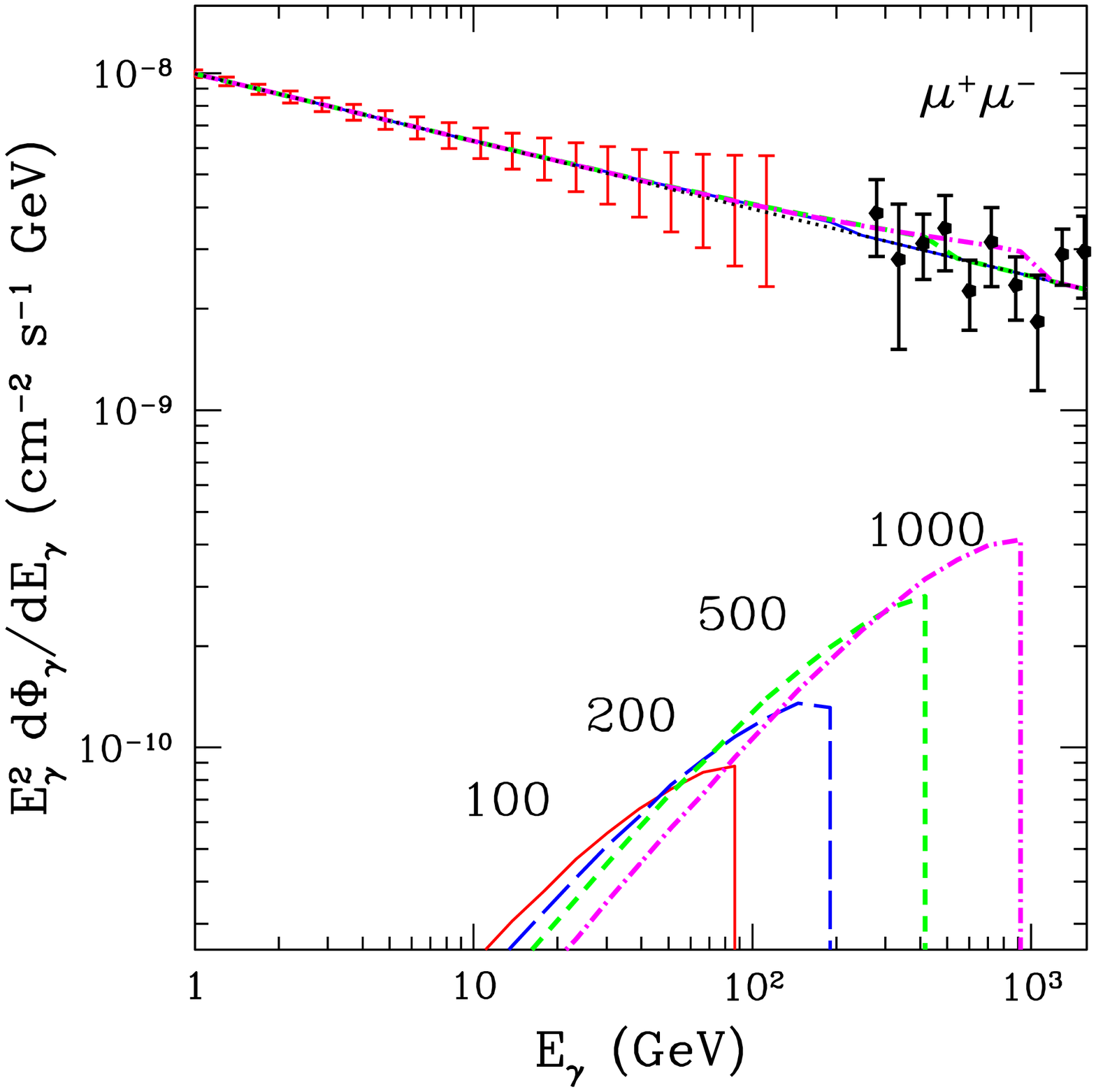}}
\resizebox{7.5cm}{!}{\includegraphics{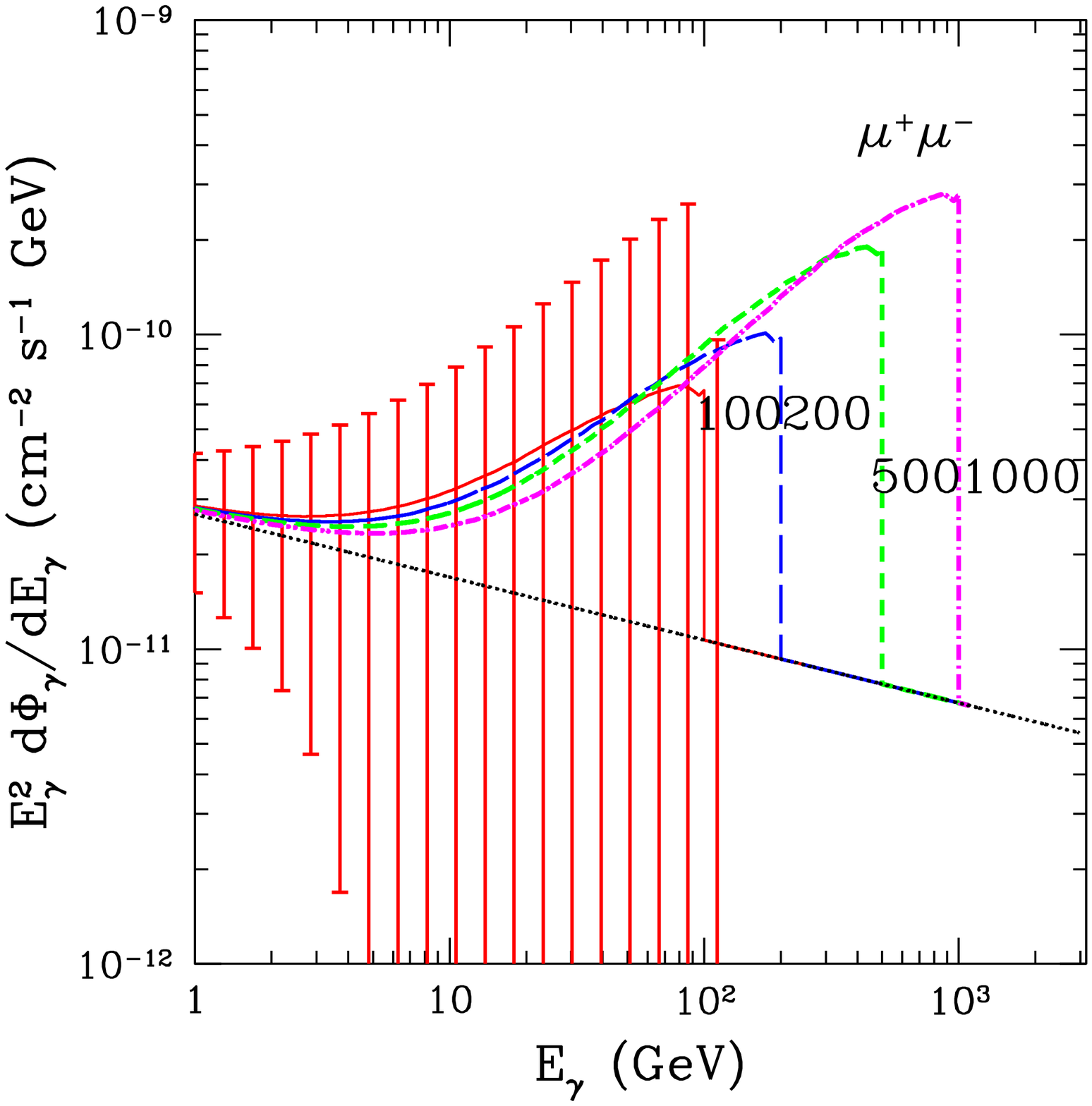}} \\
\resizebox{7.5cm}{!}{\includegraphics{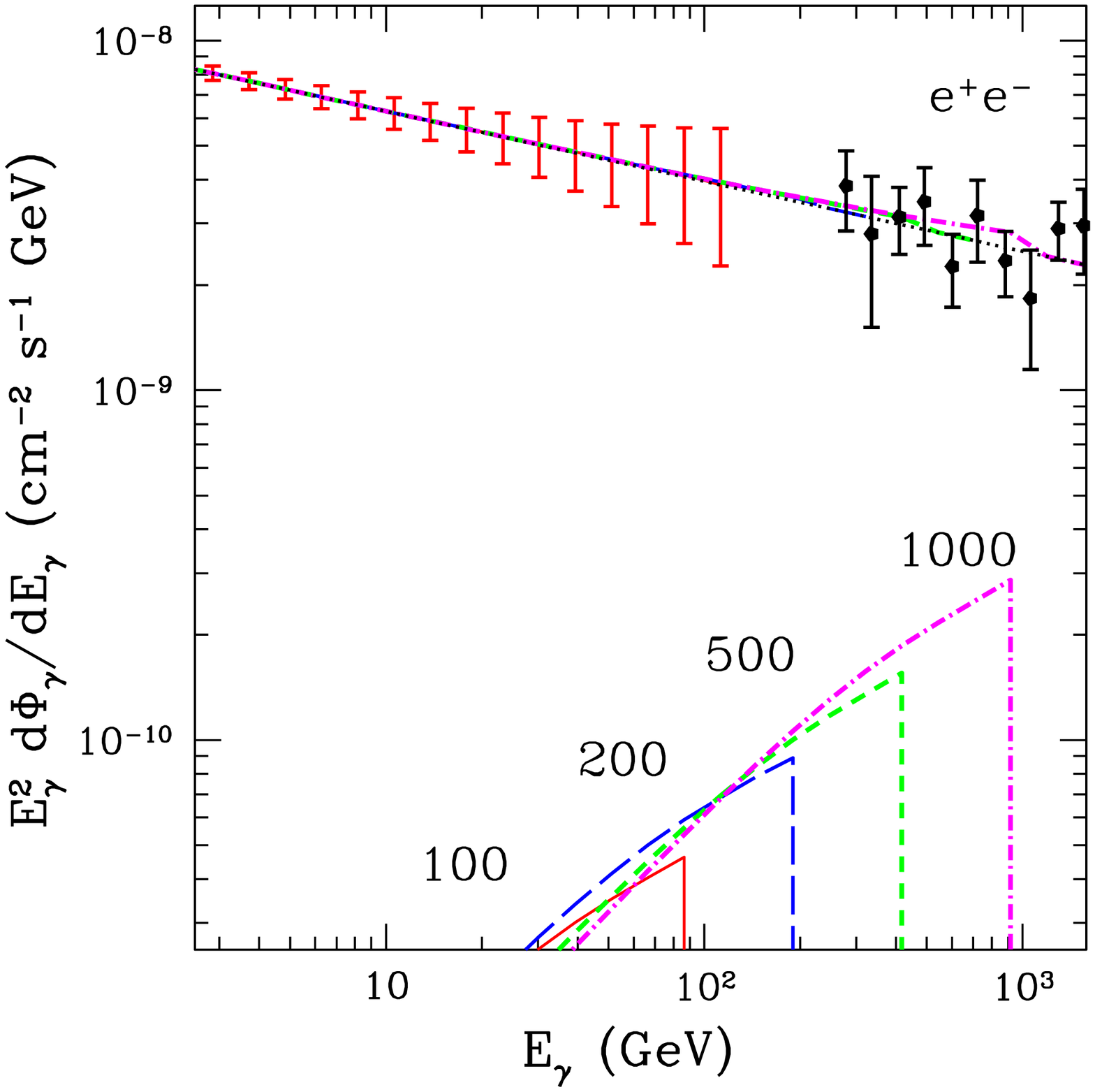}}
\resizebox{7.5cm}{!}{\includegraphics{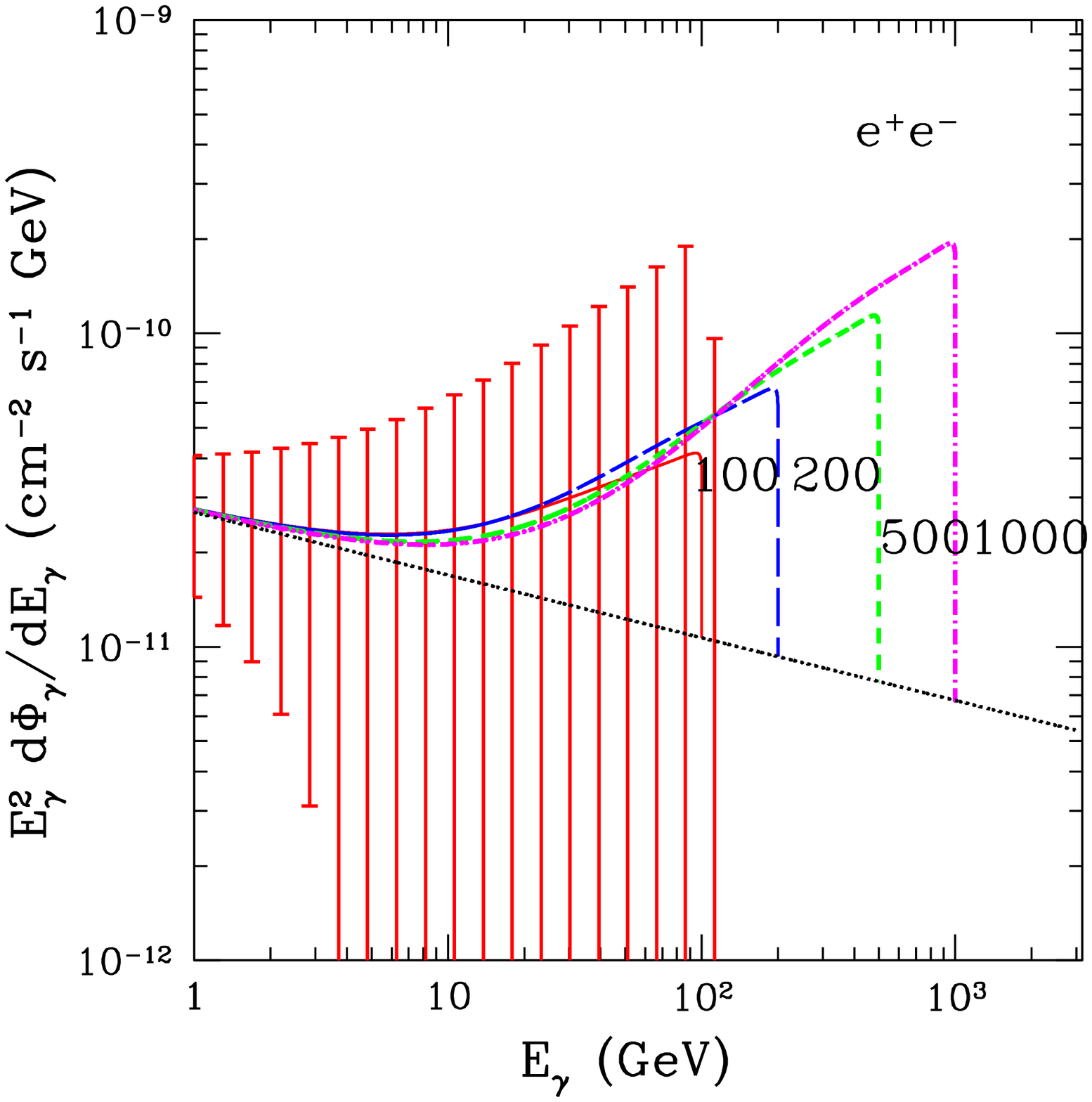}}
\caption{The same as in Fig.~\ref{gcspec}, but for dark matter annihilations to charged lepton pairs. Although the prospects for the $\tau^+ \tau^-$ mode are still promising, dark matter annihilations to $\mu^+ \mu^-$ or $e^+ e^-$ are unlikely to be identified by GLAST.}
\label{gcspec2}
%\end{center}
\end{figure}

In figures~\ref{gcspec} and~\ref{gcspec2} we plot the gamma ray spectrum from dark matter annihilations near the Galactic Center for several WIMP annihilation modes and masses. In each of the left frames, we have considered observations averaged over a solid angle of $10^{-5}$ sr centered around the Galactic Center. In the right frames, we have considered the spectrum observed from the ring between 0.3 and 0.5$^{\circ}$ away from the Galactic Center. In each case, the results reflect the point spread function of GLAST. In the right portion of each of the left frames, we show the current measurements (error bars) from the HESS experiment and, as a dotted line, the extrapolated power-law spectrum corresponding to this astrophysical background. The dark matter annihilation spectra are shown separately and combined with this background. Error bars projected for five years of GLAST operation are also shown for the lightest WIMP's mass in each frame.

From these results, it is clear that significant departures from the background power-law spectrum could be identified by GLAST in many of the cases shown. In particular, relatively light WIMPs annihilating mostly to gauge bosons or bottom quarks generate especially distinctive signatures. Even if only the observations of the inner $10^{-5}$ sr are considered, we find that WIMPs annihilating largely to gauge bosons or heavy quarks will be identifiable by GLAST with high precision (5$\sigma$) if they are lighter than approximately 320-500 GeV and be detected with 3$\sigma$ or greater significance if lighter than 500-750 GeV. If observations at angles away from the Galactic Center are also considered, much heavier WIMPs can also be identified. Annihilations to tau pairs are somewhat more difficult to observe, but are still well within the reach of GLAST. Annihilations to muons or electrons are, in contrast, unlikely to be detected by GLAST. The reasons for this conclusion are two fold. Firstly, annihilations to light leptons generate fewer gamma rays than other modes, especially at low to moderate energies (see Fig.~\ref{spectra}). Additionally, as can be seen from Fig.~\ref{sigma}, annihilations to light leptons produce a larger flux of energetic electrons and positrons, and therefore require a smaller annihilation cross section to produce the observed intensity of the WMAP Haze.

%\begin{figure}
%%\begin{center}
%\resizebox{7.5cm}{!}{\includegraphics{chisq_gc.eps}}
%\caption{The ability of GLAST to detect a departure from the power-law spectrum of the HESS source at the Galactic Center from a dark matter annihilation component, as a function of the WIMP mass and for several choices of the dominant annihilation modes. These results are for five years of observations over a solid angle of $10^{-5}$ sr (a circle of approximately 0.1$^{\circ}$ radius). The annihilation rate and halo profile have again been set by the observed features of the WMAP Haze. For WIMPs annihilating to gauge bosons or heavy quarks, their presence can be observed by GLAST with high statistical significance for masses below approximately 400-600 GeV.}
%\label{chisqgc}
%%\end{center}
%\end{figure}

\section{Caveats}
\label{caveats}

There are a number of astrophysical assumptions which we have adopted throughout our calculations. Firstly, we have adopted the dark matter halo profile described by Eq.~\ref{1pt2}, which was found in Ref.~\cite{newhaze} to fit the angular distribution of the WMAP Haze. As we have seen, the dark matter distribution in the most inner Galaxy (within $\sim 100$ parsecs of the GC) is the most relevant for dark matter searches using gamma rays. The dark matter density in the inner parsecs is only indirectly constrained by the WMAP Haze data, which is limited to angles greater than 6$^{\circ}$ from the GC (at angles within a few degrees of the Galactic Center, the microwave foregrounds are too complicated to facilitate a robust measurement of the Haze). The signal to be observed by GLAST depends critically on the slope of the halo profile at distances from the GC far smaller than the 900 parsecs which corresponds to 6$^{\circ}$. Dark matter annihilations which occur in the inner parsecs of our galaxy, however, produce electrons and positrons which diffuse over kiloparsec scales as they generate the Haze. Therefore, at an angle of 6$^{\circ}$, the intensity of the WMAP Haze provides us with a measure of the overall annihilation rate in the inner kiloparsecs of the Milky Way. In fact, using a halo profile which scales as $\rho(r) \propto r^{-1.2}$, more than 60\% of the annihilation power in the inner two kiloparsecs comes from within 100 parsecs of the GC. Therefore, if the the slope of the halo profile were to significantly flatten in the region most important to GLAST, the angular distribution of the intensity of the WMAP Haze would also be effected.

Secondly, our results depend somewhat on the treatment of the propagation of electrons and positrons in the inner Galaxy. In Ref.~\cite{newhaze}, the diffusion-loss equation was solved using specific values of the diffusion constant, energy loss rate and diffusion zone boundary conditions: $K(E_e) = 10^{28} (E_e/\rm{GeV})^{0.33}$ cm$^2$/s, $b(E_e)=5\times 10^{16} (E_e/\rm{GeV})^2$ s$^{-1}$ and a slab of half-width 3 kiloparsecs, respectively (See also Ref.~\cite{prop}). As the characteristics of the magnetic and radiation fields in the inner Galaxy are not well known, significant variation of these parameters from these values is certainly possible. 

Variations in the diffusion parameters could potentially modify the angular distribution, spectrum and intensity of the synchrotron emission from dark matter annihilation products. Although it is somewhat difficult to estimate the range of values these parameters could plausibly take, it seems highly unlikely that the true values are more than a factor of a few different from those used here. If both the diffusion constant and energy loss rate were increased by a factor of a few, it would have the same effect of reducing the annihilation rate by the same factor. This would unlikely change the overall conclusions of our study. 

Thirdly, astrophysical gamma ray backgrounds other than the HESS source discussed earlier are likely to exist in the inner degree surrounding the Galactic Center. In particular, an unidentified EGRET source is known to be present approximately 0.2$^{\circ}$ from the Galactic Center~\cite{dingus}. The angular windows around such point sources could be subtracted from the GLAST spectrum, however. Alternatively, an energy cut could be made on the GLAST data to reduce the impact of such backgrounds.

\section{Summary and Conclusions}

If the WMAP Haze is, in fact, synchrotron emission from relativistic electrons and positrons produced in dark matter annihilations, then other annihilation products are likely to be potentially observable as well. In particular, gamma rays from dark matter annihilations in the inner parsecs of the Milky Way are potentially within the reach of the upcoming satellite-based gamma ray telescope GLAST. 

In this article, we study the prospects for GLAST to detect gamma rays from dark matter annihilations in the inner Galaxy, assuming that such annihilations are also responsible for the observed microwave excess known as the WMAP Haze. We find that if the dark matter particles annihilate largely to heavy fermions, or gauge or Higgs bosons, then their gamma ray spectrum should be easily identified by GLAST. Only if dark matter annihilations proceed largely to muons or electrons will such a detection not be possible.

It is also interesting to note that the Planck satellite is expected to begin its mission in 2008. With its much broader frequency coverage, Planck will be capable of determining the properties of the WMAP Haze with much greater precision. The combined scrutiny of GLAST and Planck should clearly resolve the nature of the Haze and determine whether it is of dark matter origin once and for all.

\smallskip

DH is supported by the US Department of Energy and by NASA grant NAG5-10842. DF and GB are partially supported by NASA grant NAGS-12972. Discussions of the inverse Compton signal with Nikhil Padmanabhan were helpful.

\end{document}